\documentclass{PoS}

\title{Systematic Effects at Criticality for the SU(2)-Landau-Gauge 
Gluon Propagator}

\ShortTitle{Systematic Effects at Criticality for the Gluon Propagator}

\author{\speaker{Tereza Mendes}\\
        Instituto de F\'\i sica de S\~ao Carlos, Universidade de S\~ao Paulo,\\
        Caixa Postal 369, 13560-970 S\~ao Carlos, SP, Brazil\\
        E-mail: \email{mendes@ifsc.usp.br}}

\author{Attilio Cucchieri\\
        Instituto de F\'\i sica de S\~ao Carlos, Universidade de S\~ao Paulo,\\
        Caixa Postal 369, 13560-970 S\~ao Carlos, SP, Brazil\\
        E-mail: \email{attilio@ifsc.usp.br}}

\abstract{
We analyze data from finite-temperature simulations of the gluon propagator
in SU(2) Landau gauge on large lattices. We argue that the singular behavior
of this quantity around the deconfinement transition, seen in several 
previous studies, is a lattice artifact.
         }

\FullConference{31st International Symposium on Lattice Field Theory - LATTICE 2013\\
		July 29 - August 3, 2013\\
		Mainz, Germany}

\begin{document}

\section{Introduction}
The gluon propagator is the most fundamental quantity of QCD
and its infrared behavior is believed to be closely related
to the phenomenon of color confinement in the theory. In particular,
the Gribov-Zwanziger confinement scenario in Landau gauge 
\cite{Vandersickel:2012tz} predicts a suppressed gluon propagator 
in the infrared limit (in combination with an enhanced ghost propagator).
According to this scenario, in fact, the gluon propagator should go to zero
in the limit of vanishing momentum.
These predictions are investigated by approximate analytic methods such as
Dyson-Schwinger equations and functional renormalization group
calculations. At the same time, the lattice formulation can provide 
valuable insight into the problem and numerical checks of the predictions. 
Unfortunately, the 
infrared limit corresponds to large lattice sizes, which are 
computationally demanding. This issue has turned out to be particularly 
challenging in Landau gauge, requiring numerical investigations using
the largest lattices ever considered (see \cite{Cucchieri:2010xr} for a review).
Nevertheless, the infrared limit may be qualitatively studied for pure
SU(2) gauge theory and, at the same time, using very large lattices might
greatly reduce the infamous problem of gauge-fixing ambiguity due to 
Gribov copies \cite{Zwanziger:2003cf}.

\begin{figure}[t]
\ \hskip -14mm \ 
\includegraphics[height=6.8truecm]{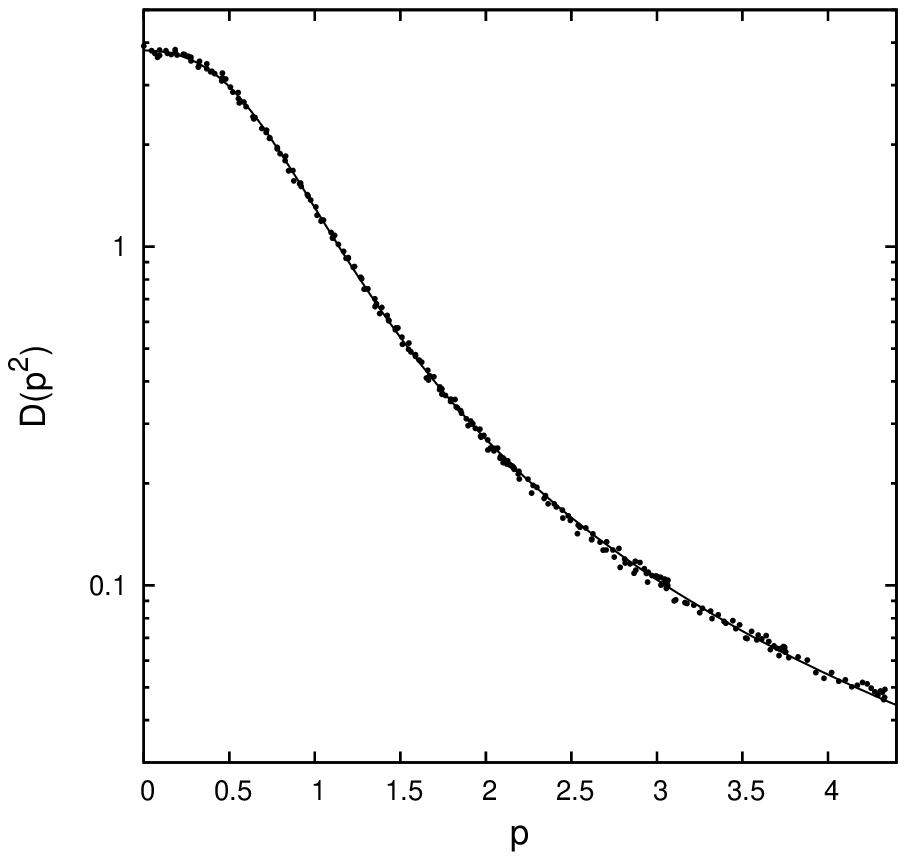}
\ \hskip -20mm \ 
\includegraphics[height=6.8truecm]{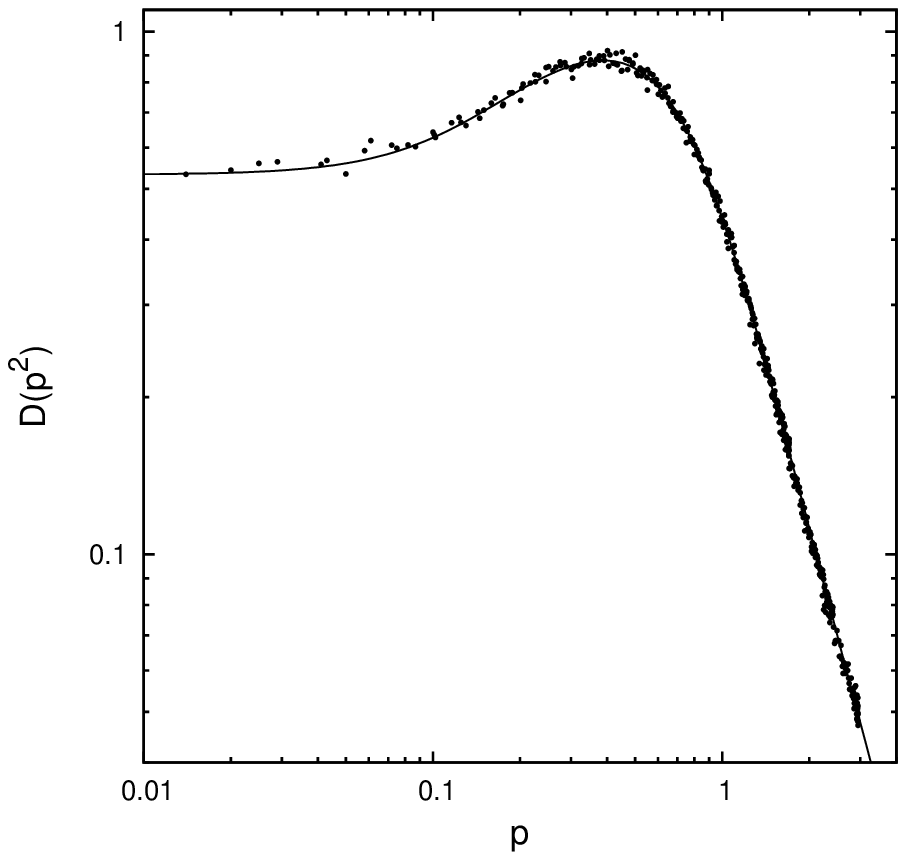}
\caption{Fits of zero-temperature data for the SU(2) Landau-gauge
gluon propagator to rational (Gribov-Stingl) forms in the 4D (left) 
and 3D (right) cases.
Plots extracted from \cite{Cucchieri:2011ig}.}
\label{fig:fits}
\end{figure}

Lattice simulations have established that the momentum-space gluon 
propagator $D(p^2)$ is suppressed in the limit of small momentum $p$, while 
the real-space gluon propagator violates reflection positivity. This latter 
feature, consistent with gluon confinement, is observed for all lattice 
volumes. On the other hand, whereas a fit of $D(p^2)$ to the Gribov form
is possible at moderate lattice volumes, data obtained using very large 
lattices (of linear size $L \approx 27$ fm) revealed that $\,D(0)\,$ 
is strictly nonzero.
This behavior has been termed ``massive'', since it may be interpreted
as a dynamically generated mass for the gluon, and was first proposed
as a solution to the Dyson-Schwinger equations of QCD in \cite{Cornwall:1981zr}.
Several variants of such massive behavior have been used to fit lattice
data for the Landau-gauge gluon propagator.
In particular, in \cite{Cucchieri:2011ig}, very good fits to rational (or
Gribov-Stingl) forms were obtained in the four-dimensional case, as well
as for three space-time dimensions.
These fits are shown in Fig.\ \ref{fig:fits}.
The fitting forms in the 4D and in the 3D cases are given respectively by
\begin{equation}
D_1(p^2) \;=\; C\;\frac{p^{2}\,+\,d}
{p^4 \,+\, u^2\,p^2 \,+\, t^2}\,
\label{eq:GS1}
\end{equation}
and
\begin{equation}
D_2(p^2) \;=\;  C\;\frac{(p^{2}\,+\,d)\,(p^{2}\,+\,1)}
{(p^4 \,+\, u^2\,p^2 \,+\, t^2)\,(p^{2}\,+\,v)}\,,
\label{eq:GS2}
\end{equation}
corresponding (respectively) to three and to four free parameters,
in addition to the global normalization constant $C$.
Noting that the three-dimensional case may be viewed as the 
infinite-temperature limit of the four-dimensional case in the 
transverse sector, one may be motivated to look for an interpolation
of the above 4D and 3D zero-temperature forms to describe the
finite-temperature data for the propagator.

In this contribution we present final results of our numerical 
study of the finite-temperature gluon propagator in the electric sector.
We focus on the infrared value of the longitudinal propagator $D_L(p^2)$
as a function of the temperature.
A detailed analysis of our data will be presented elsewhere \cite{inprep}. 
(Preliminary results were reported in
\cite{Cucchieri:2011ga,Cucchieri:2011di,Cucchieri:2012nx,Cucchieri:2012gb}.)


\section{Gluon propagator at finite temperature}

As the temperature $T$ is turned on, we expect to observe 
Debye screening of the color charge. In particular, at
high temperatures, deconfinement should be felt in the 
longitudinal (i.e.\ electric) gluon propagator
as an exponential fall-off at long distances,
defining a screening length and conversely a screening mass
\cite{Gross:1980br}.
It is not clear how such a mass would show up around the critical 
temperature $T_c$.
At the same time, as discussed above, studies of the gluon propagator
at zero-temperature have shown a (dynamical) mass. One can try to 
use this knowledge to define temperature-dependent masses for the 
region around $T_c$.
Conversely, the dimensional-reduction picture
(based on the 3D-adjoint-Higgs model) suggests a
confined magnetic gluon, associated to a nontrivial magnetic mass.
This mass should in turn be obtained from the infrared behavior of
the transverse gluon propagator.

Lattice studies of the Landau-gauge gluon propagator around the deconfinement
phase transition in pure $SU(2)$ and $SU(3)$ theory, as well as considering
dynamical quarks, have been presented in \cite{Cucchieri:2007ta,Fischer:2010fx,Bornyakov:2010nc,Aouane:2011fv,Maas:2011ez,Bornyakov:2011jm,Aouane:2012bk,Silva:2013maa}.
In the transverse (i.e.\ magnetic) sector, one sees strong infrared 
suppression of the propagator, with a turning point of the curve described by 
the momentum-space magnetic propagator $D_T(p^2)$ for momenta $p$ around 
400 MeV. This suppression seems even more pronounced than 
in the zero-temperature case discussed in the Introduction. 
Also, $D_T(p^2)$ shows considerable finite-physical-size 
effects in the infrared limit, as observed for $T=0$. 
Furthermore, just as for $T=0$, the magnetic propagator 
displays a clear violation of reflection positivity in real space.
Essentially these same features are seen for $D_T(p^2)$ at all nonzero
temperatures considered.

The longitudinal propagator $D_L(p^2)$, on the other
hand, shows significantly different behavior for different temperatures.
As soon as a nonzero temperature is introduced in the system,
$D_L(p^2)$ increases considerably (whereas $D_T(p^2)$ decreases
monotonically). More precisely, for all fixed temperatures, the curve 
described by $D_L(p^2)$ seems to reach a plateau in the low-momentum region
(see e.g.\ \cite{Cucchieri:2011di}). As the temperature is increased,
this plateau increases slightly until, approaching the phase transition
from below, it has been observed to rise further and then, just above
the transition temperature, to drop sharply. This has been interpreted
as a sign of singular behavior of the longitudinal gluon propagator 
around $T_c$ and, in fact, it has been related to several proposals of a
new order parameter for the deconfinement transition.
(Of course, a relevant question is, then, whether this singularity survives
the inclusion of dynamical quarks in the theory 
\cite{Bornyakov:2011jm,Aouane:2012bk}.)

Let us mention that, at all investigated temperatures, the infrared 
plateau just described is not long enough to justify a fit to the 
Yukawa form 
\begin{equation}
D_{L}(p^2) \;=\;
C\,\frac{1}
{p^2 \,+\, m^2}\,,
\label{eq:Yukawa}
\end{equation}
predicted at high temperatures. If this were the case, $D_L(0)^{-1/2}$
would provide a natural (tem\-pe\-ra\-tu\-re-dependent) mass scale. 
Note that this value depends also on the global constant $C$. 
On the other
hand, Gribov-Stingl forms such as in Eqs.\ (\ref{eq:GS1}) and (\ref{eq:GS2}) 
above involve complex-conjugate poles, defining real and imaginary masses
(independently of $C$).
Here we do not show data (or fits) for $D_L(p^2)$.
Such curves and (preliminary) fits can be seen e.g.\ in 
\cite{Cucchieri:2012nx}.
Instead, we will look at the value of $D_L(0)$ 
(after normalization by $C$) as a function of $T$.

Concerning the longitudinal propagator in real space 
(see e.g.\ \cite{Cucchieri:2012nx}), positivity violation is observed 
unequivocally only at zero temperature
and for a few cases around the critical region, in association with
the severe systematic errors discussed below. For all other cases,
there is no violation within errors.
Also, we always observe an
oscillatory behavior, indicative of a complex-mass pole.
In the next section, we present our new results for the infrared
values of $D_L(p^2)$.


\section{Results}

Our large-lattice study was done considering the pure SU(2) case, 
with a standard Wilson action and lattice sizes $\,N_s^3 \times N_t$
ranging from $48^3 \times 4$ to $192^3 \times 16$. 
For our runs we employ a cold start, performing a projection on
positive-Polyakov-loop configurations.
Also, gauge fixing is implemented using stochastic overrelaxation.
The gluon dressing functions are normalized to 1 at 2 GeV.
We considered several values of the lattice parameter $\beta$, allowing 
a broad range of temperatures. 
Our procedure for determining the physical temperature $T$
is described in \cite{Cucchieri:2011di}.
The momentum-space expressions for the transverse and longitudinal
gluon propagators $D_T(p^2)$ and $D_L(p^2)$ can be found e.g.\
in \cite{Cucchieri:2007ta}.

As can be seen from the data in \cite{Cucchieri:2012nx},
the longitudinal (electric) propagator $D_L(p^2)$ displays severe
systematic effects around $T_c$ for the smaller values of $N_t$.
These effects are strongest at temporal extent $N_t=4$ and large
values of $N_s$. We note that the systematic errors for small
$N_t$ come from two different sources: ``pure'' small-$N_t$ effects 
(associated with discretization errors) and strong dependence on the
spatial lattice size $N_s$ at fixed $N_t$, for the cases in which
the value of $N_t$ is smaller than 16. The latter effect was
observed only at $T \lesssim T_c$, whereas the former is present in a wider 
range of temperatures around $T_c$.
In particular, the finite-spatial-volume effects for $D_L(p^2)$ 
at $N_t = 4$ are strongest at $T_c$, but are still very large at 
$T=0.98 T_c$ and are much less pronounced for $T=1.01 T_c$.

\begin{figure}
\ \hskip -15mm \ 
\includegraphics[height=12truecm]{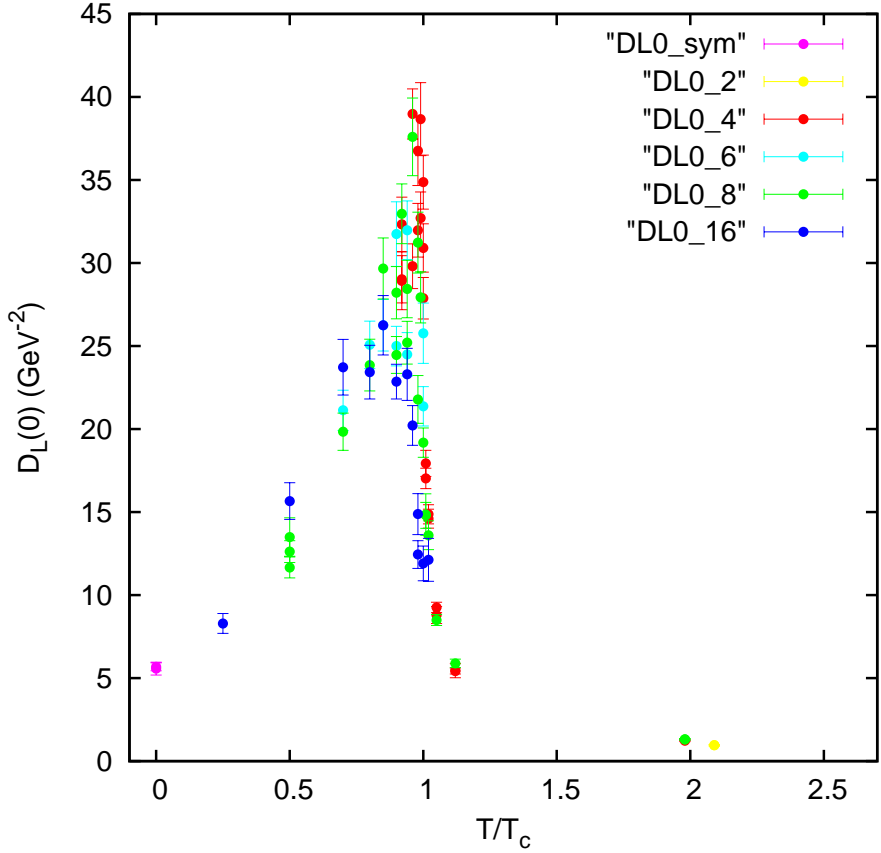}
\caption{Infrared-plateau value for the longitudinal gluon propagator 
[estimated by $D_L(0)$] as a function of the temperature for the full 
range of $T/T_c$ values.
Data points from runs at the same value of $N_t$ are grouped together and 
indicated by the label ``DL0\_${N_t}$'', where ``sym'' is used
to indicate symmetric lattices (i.e.\ $T=0$).}
\label{fig:bump}
\end{figure}
\begin{figure}
\ \hskip -15mm \ 
\includegraphics[height=12truecm]{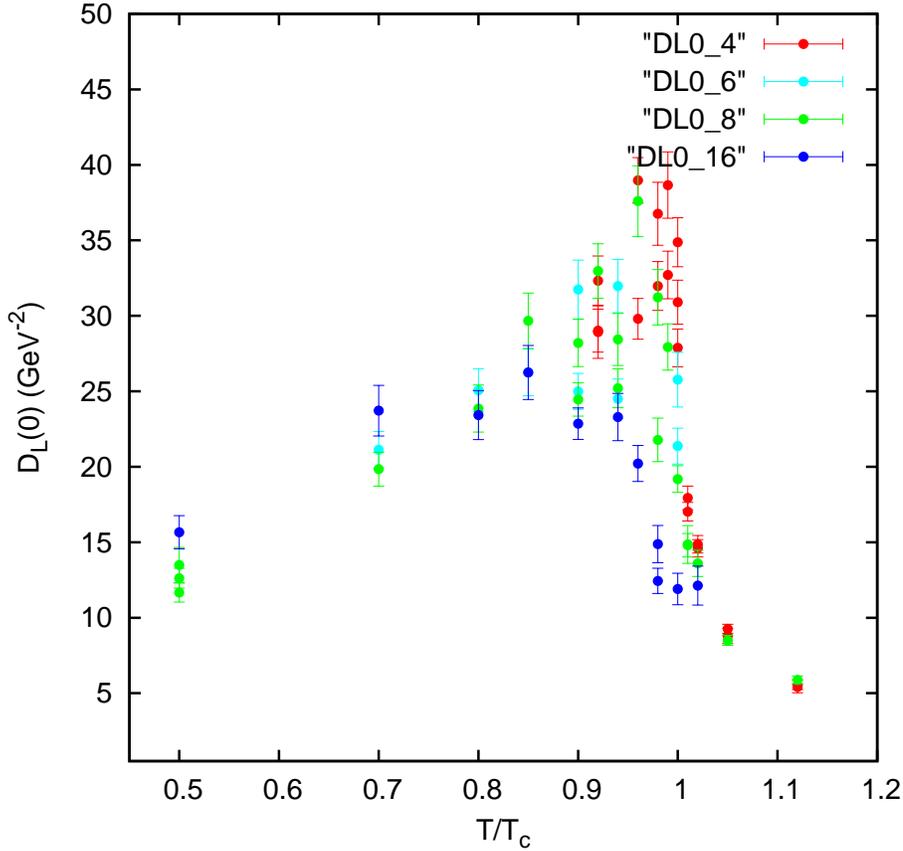}
\caption{Same as Fig.\ \protect\ref{fig:bump} above, but showing only the
temperature region around $T_c$.}
\label{fig:bump_zoom}
\end{figure}

In Fig.\ \ref{fig:bump} we show data for $D_L(0)$ as a function of the
temperature $T$. We show such values as obtained from all our runs, 
grouping together (by color) the runs performed at the same temporal 
extent $N_t$. We remark that, as said above, not all curves of $D_L(p^2)$ 
reach a clear plateau in the infrared limit. Nevertheless, looking at the
value of $D_L(0)$ gives us an indication of what this plateau might be,
and is useful to expose the strong systematic effects discussed here.
In Fig.\ \ref{fig:bump_zoom} we present an enlarged view of the same
values for the temperature region around $T_c$.

We can see that the very suggestive sharp peak at $T_c$ seen for $N_t=4$
(corresponding to the red points in Figs.\ \ref{fig:bump} and
\ref{fig:bump_zoom})
turns into a finite maximum around 0.9 $T_c$ for $N_t=16$ (blue points).
In other words, the observed 
singularity at smaller values of $N_t$ seems to disappear. The only
indication of a possible singular behavior is a finite maximum close to 
(but not {\em at}) the critical point, somewhat reminiscent of a 
pseudo-critical point
as observed for the magnetic susceptibility of spin models in an
external magnetic field (see e.g.\ \cite{Cucchieri:2004xg,Engels:2011km}).

Let us mention that, as reported in \cite{Cucchieri:2012nx}, good fits 
are obtained (in the transverse and longitudinal cases) to several 
generalized Gribov-Stingl forms,
indicating the presence of comparable real and imaginary parts of pole masses.
These masses are smooth functions of $T$ around the transition, and the
imaginary part of the electric mass seems to get smaller at higher $T$, as
expected.


\section{Conclusions}

We have performed numerical simulations of the longitudinal (electric) 
and transverse (magnetic) Landau-gauge gluon propagator at nonzero
temperature for pure SU(2) lattice gauge theory. We employ the largest
lattices to date, especially for temperatures around the deconfinement
phase transition. We are currently completing our study of fitting forms 
for describing the massive behavior of the propagator \cite{inprep}.
From our data for the longitudinal gluon propagator $D_L(p^2)$, we have
uncovered quite severe systematic effects.

Our results point to unusually large systematic errors around 
criticality. In particular, very strong effects related to small values
of the temporal extent $N_t$ of the lattice are seen on the lower 
side of the transition temperature and are practically absent just
above $T_c$.
Strong finite-size effects are certainly not unexpected around a
second-order phase transition, such as the deconfinement transition
in the SU(2) theory. On the other hand, we note that our data show a 
nontrivial dependence on the finite {\em temporal} size of the lattice 
and on the distance from the critical point, not easily interpreted as a 
finite-size or a discretization effect. 

After removing these systematic effects, i.e.\ considering the data obtained
with the largest value of $N_t$ in Fig.\ \ref{fig:bump_zoom}, we see that
the sharp peak suggested by the red points in Fig.\ \ref{fig:bump} turns 
into a smooth maximum, at around $0.9\,T_c$.
In agreement with several observations that the gluon mass scale is a smooth
function of the temperature, this suggests that there is no specific signature 
of deconfinement associated with $D_L(p^2)$.
In fact, the only qualitative feature of a deconfined phase we observe
is the lack of violation of reflection positivity for the real-space 
electric propagator, which holds however for all $T\neq 0$ considered.

Finally, let us mention the similarity between our smaller-lattice results 
for the SU(2) case and existing results for SU(3).
In view of this, we trust that statements that the inverse of the
zero-momentum value of the gluon propagator might provide an order parameter
for the deconfinement phase transition (such as recently made in
\cite{Silva:2013maa}) will be taken with the due caution.


\end{document}